\documentclass{PoS}

\title{Heavy Flavours in Quark-Gluon Plasma}

\ShortTitle{Heavy Flavors in QGP}

\author{\speaker{Seyong Kim}\\ %\thanks{A footnote may follow.}\\
        Department of Physics, Sejong University, Seoul 05000, Republic of Korea\\
        E-mail: \email{skim@sejong.ac.kr}}

\abstract{Recent progresses in lattice studies of heavy quark and
  quarkonium at non-zero temperature are discussed. Formulating a tail
  of spectral functions as a transport coefficient allows lattice
  determination of momentum diffusion coefficient ($\kappa$) for charm
  quark in the heavy quark mass limit and lattice determination of
  heavy quark/heavy anti-quark chemical equilibration rate in
  NRQCD. Quenched lattice study on a large volume gives $\kappa/T^3 =
  1.8 \cdots 3.4$ in the continuum limit. A recent study with $N_f =
  2+1$ configurations estimates the charmonium chemical equilibration
  rate $\Gamma_{\rm chem}$. At $T = 400$ MeV with $M \sim 1.5$ GeV,
  $\Gamma_{\rm chem}^{-1} \sim 150$ fm/c. Earlier results from the two
  studies (with different lattice setups and with different Bayesian
  priors) which calculate bottomonium correlators using NRQCD and
  employ Bayesian method to calculate spectral functions are
  summarized: $\Upsilon (1S)$ survives upto $T \sim 1.9 T_c$ and
  excited states of $\Upsilon$ are sequentially suppressed. The
  spectral functions of $\chi_{b1}$ channel shows a Bayesian prior
  dependence of its thermal behavior: the $\chi_{b1}$ spectral
  function with MEM prior shows melting above $T_c$ but that with a
  new Bayesian prior hints survival of $\chi_{b1}$ upto $\sim 1.6
  T_c$. Preliminary results from the efforts to understand the
  difference in the behavior of $\chi_{b1}$ spectral function is
  given.  }

\FullConference{34th annual International Symposium on Lattice Field Theory\\
		24-30 July 2016\\
		University of Southampton, UK}

\begin{document}

\section{Introduction}

% experimental motivation and advantages  and difficulties of lattice
% calculation 
%
% (1) zero temperature situation
% (2) non-zero temperature situation
% (3) what lattice can do
% (4) obstacles and solutions for lattice calculations
%  

Understanding the properties of Quark-Gluon Plasma (QGP) and the
transition from the hadronic gas phase to the QGP phase quantitatively
is one of the most important subjects in the studies of QCD
thermodynamics. Heavy flavours (charm and bottom quark) open an
important window for the lattice community to this question together
with accompanying difficulties. 

\begin{figure}[ht]
\centering
\begin{minipage}[b]{0.55\linewidth}
\includegraphics[width=\textwidth]{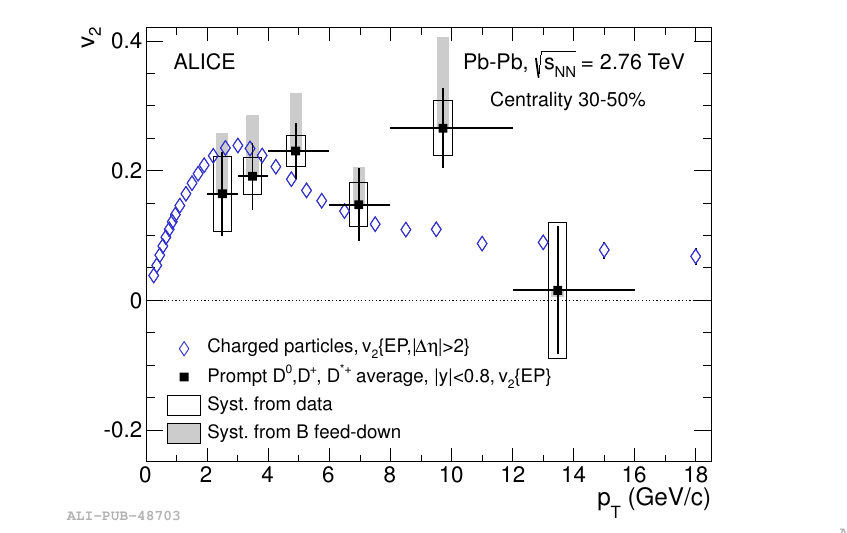}
\end{minipage}
\quad
\begin{minipage}[b]{0.35\linewidth}
\includegraphics[width=\textwidth]{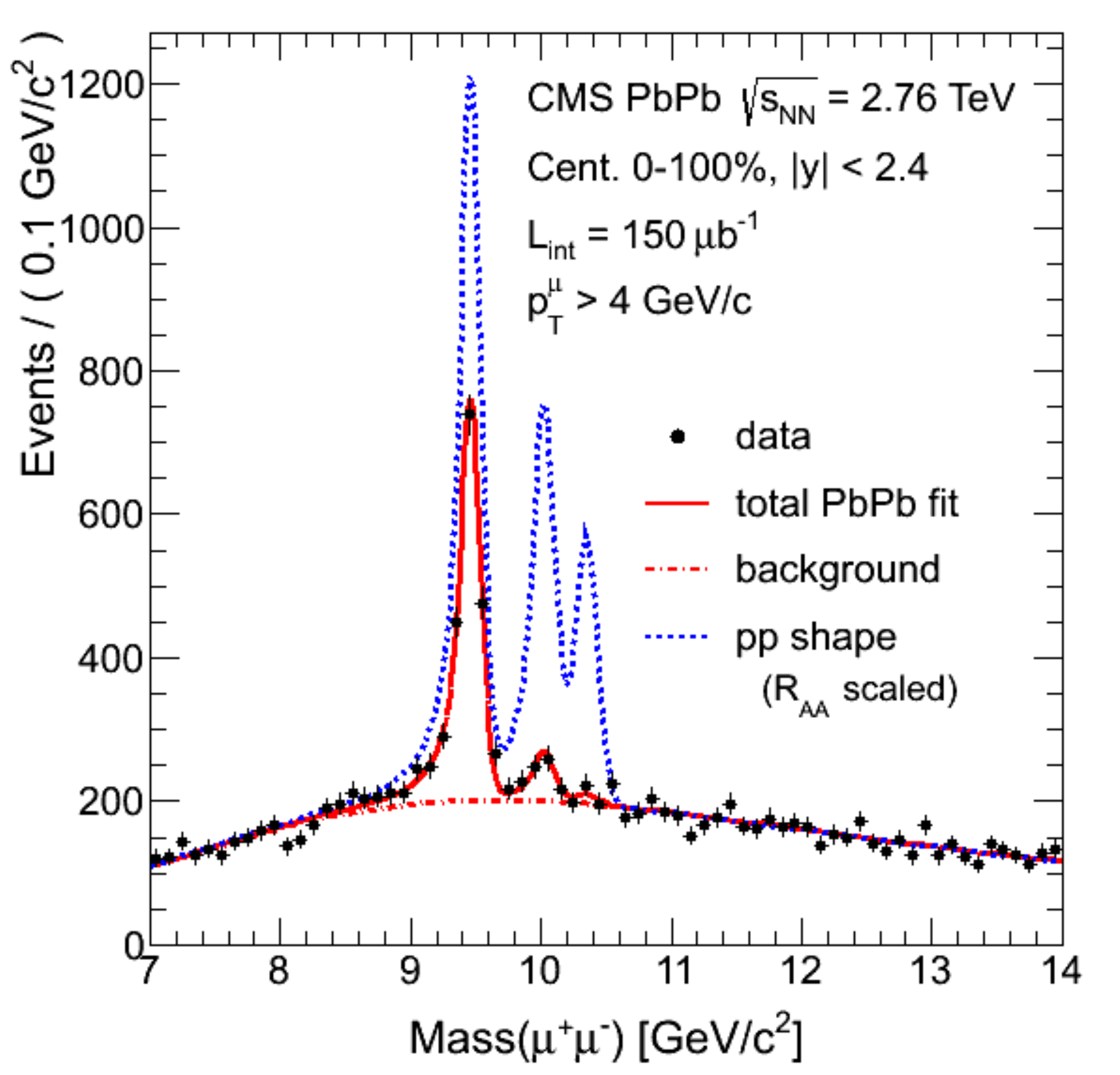}
\end{minipage}
\caption{Elliptic flow ($v_2$) of light charged hadron and charmed meson
  ($D^0, D^+, D^{*+}$) from ALICE collaboration
  \cite{Abelev:2014ipa} (left) and sequential suppression of
  $\Upsilon (1S), \Upsilon (2S)$ and $\Upsilon (3S)$ from CMS
  collaboration \cite{Chatrchyan:2012lxa}.} 
\label{fig:Exp_ALICE_CMS}
\end{figure}

Experimentally, investigations of QGP properties usually require
comparisons between results of proton-proton collisions and those of
relativistic heavy ion collisions (see e.g., the right figure in
Fig. \ref{fig:Exp_ALICE_CMS} where $\Upsilon$ production in $p-p$
collisions is compared with that in $Pb-Pb$ collisions) and having
better grasps on the baseline properties of $p-p$ collisions help us
to better distinguish the differences between quarkonium in
non-interacting collections of hadron-hadron collisions and that in
nucleus-nucleus collisions. In this regard, hadronic processes which
involve heavy quark(s) can be advantageous since these processes can
be understood more precisely due to ``factorization'' of the
corresponding processes in terms of the short distance perturbative
interactions and the long distance non-perturbative matrix elements
through effective field theories such as Non-Relativistic QCD
\cite{Bodwin:1994jh}, potential NRQCD
\cite{Brambilla:2004jw}. The same effective field theories can be
applied in non-zero temperature environment as long as $T/M << 1$.

Also, as illustrated by the recent measurement of large elliptic flow
of charmed meson (the left figure in Fig. \ref{fig:Exp_ALICE_CMS})
from ALICE collaboration, behaviors of heavy quarks in relativistic
heavy ion collisions which may be viewed as ``a heavy quark in thermal
medium'' have more than a few surprises and raise many interesting
questions for their own sakes. Another effective theory, Heavy Quark
Effective Theory (HQET) is useful, where heavy meson (a heavy
quark-light quark bound state) can be understood as ``brown mug''
surrounding a heavy quark \cite{Isgur:1989vq} due to the heavy quark
mass scale. Recently, heavy quark mass limit in non-zero temperature
setup allowed non-perturbative formulation of otherwise difficult real
time quantities in non-zero temperature \cite{CaronHuot:2009uh,
  Bodeker:2012gs, Bodeker:2012zm, Kim:2016zyy}.

Of course, relating experiments which includes complex time evolution
of highly relativistic scattering of nuclei to the quantities which
lattice gauge theory can calculate is highly nontrivial and is
involved since lattice gauge theory mostly describes physics in
thermal equilibrium. Furthermore, studying properties of heavy quarks
and quarkonia at $T \neq 0$ temperature on a lattice faces many
challenges not only because simulations with dynamical charm quark
just become possible and simulations with dynamical bottom quark is
not possible yet: spectral behaviors of the correlators at non-zero
temperature are far more complicated than those at zero temperature
due to thermal broadening, transport phenomena, disappearance of bound
states, and etc. On the other hand, in isotropic lattice formulation,
the temporal correlators at $T \neq 0$ typically have less number of
temporal lattice sites than those at zero temperature. Deducing more
information from less data points is unavoidable.

Here, we discuss heavy quarks and quarkonia at $T \neq 0$ focusing on
(1)two real time quantities concerned with heavy quark
kinetic/chemical equilibriation and (2)in-medium quarkonium studied
with NRQCD correlators and Bayesian reconstructions of corresponding
spectral functions. Although in this conference, a talk on
``open charm on anisotropic lattice'' \cite{Skullerud}, a talk on new
attempt on the inverse integral transform method together with a new
determination of charm diffusion constant \cite{Ikeda}, another talk
on new determination of the charmonium diffusion constant \cite{Ohno},
a talk on new stochastic method for reconstructing spectral function
\cite{Shu}, and a talk on the Debye screening mass of the potential
determined by first principles method \cite{Rothkopf} are given, due
to lack of space, these are not covered by this plenary talk.

\section{Heavy Quarks}

Hydrodynamic flow of charmed meson in relativistic $Pb-Pb$ collisions
is comparable to that of light hadron (Fig. \ref{fig:Exp_ALICE_CMS}),
which suggests that an effective thermalization of charm quark is
similar to that of light quarks \cite{Abelev:2014ipa}. This is in
conflict with a perturbative consideration and calls for a lattice QCD
determination of transport properties of heavy quarks in QGP.
Transport properties are non-equilibrium characteristics and are
usually accessed by the ``transport peak'' of the spectral function
for the corresponding correlators in thermal equilibrium through Kubo
relations \cite{Kapusta:2006pm}. However, obtaining the transport
coefficient directly from the ``transport peak'' of the spectral
function for heavy quark current-current correlator is hard because
the width of transport peak in the spectral function scales as $\sim
\alpha_s^2 T^2/M$ and the peak is narrow for the heavy quark case. A
HQET based formulation by \cite{CaronHuot:2009uh} allows one to
extract a transport coefficient from the ``power-law frequency tail''
of the spectral function, instead of ``zero frequency limit peak'' of
the spectral function. A quenched lattice determination of the
transport coefficient related to the kinetic equilibriation of heavy
quark in thermal environment following this formulation is reported
recently in \cite{Francis:2015daa}

One of other interesting questions concerning the behavior of heavy
quarks in QGP is whether the abundance of heavy quarks/anti-heavy
quarks in QGP follows thermal Boltzmann distribution. Experimentally,
it is related to whether the chemical equilibriation rate of heavy
quarks is large compared to the lifetime of QGP achieved in
relativistic heavy ion collisions. Similarly to the case of kinetic
equilibriation of heavy quark in thermal medium
\cite{CaronHuot:2009uh}, chemical equilibriation of heavy
quark/heavy anti-quark in thermal equilibrium can be nonperturbatively
defined through density-density correlators in NRQCD limit
\cite{Bodeker:2012gs, Bodeker:2012zm}. A lattice determination (in
full $N_f = 2+1$ dynamical simulation) of the chemical equilibriation
rate of heavy quark/anti-heavy quark is reported for the first time
\cite{Kim:2016zyy}.

\subsection{Kinetic equilibriation}

Let us briefly describe the extraction of a transport coefficient from
power-law tail of a spectral function. Hydrodynamic property of heavy
quark moving through a thermal medium is characterized by the
diffusion coefficient, $D$, defined as 
\begin{equation}
D = \frac{1}{3 \chi_{00}} \lim_{\omega \rightarrow 0} \sum_{i=1,..3}
\frac{\rho_V^{ii} (\omega)}{\omega}, \;\;\;\;
\rho_V^{ii} (\omega) = \int_{-\infty}^\infty dt \; e^{i \omega t} \int
d^3 x \; \langle \frac{1}{2} \left[J^i (t, {\mathbf x}), J^i (0, {\mathbf
    0}) \right] \rangle
\label{DC}
\end{equation}
where $J^i = \overline{\psi} \gamma^i \psi$ ($\psi$ is the
relativistic heavy quark field) through fluctuation-dissipation
theorem. $\chi_{00}$ is a susceptibility related to the $0$-th
component of the four-current by 
\begin{equation}
\chi_{00} = \frac{1}{T} \int d^3 x \; \langle J^0
  (t, {\mathbf x}) J^0 (t, {\mathbf 0}) \rangle .
\end{equation}
The transport peak, $\omega \rightarrow 0$ limit of $\rho_V^{ii}
(\omega)$, becomes narrower as $M \rightarrow \infty$ and is difficult
to access. Instead, from the observation that in a large heavy quark
mass limit, the momentum diffusion constant($\kappa$), the drag
coefficient ($\eta_D$) and the kinetic equilibriation time ($\tau_{\rm
  kin}$) are related to $D$ by 
\begin{equation}
D = \frac{2T^2}{\kappa}, \;\;\;\; \eta_D = \frac{\kappa}{2MT},
\;\;\;\;\; \tau_{\rm kin} = \frac{1}{\eta_D} 
\label{DandetaD}
\end{equation}
with an assumption of ``narrow transport peak'' and a Lorentzian width
around the peak \cite{CaronHuot:2009uh}, authors of
\cite{Francis:2015daa} focus on the momentum diffusion
coefficient, $\kappa$. The mass dependent momentum diffusion
coefficient, $\kappa^M$, is defined as
\begin{equation}
\kappa^M = \frac{M^2_{\rm kin} \omega^2}{3 T \chi^{00}} \sum_i \frac{2
T \rho_V^{ii} (\omega)}{\omega} |_{\eta_D << |\omega| \le \omega_{UV}},
\end{equation}
where $M_{\rm kin}$ is the kinetic mass of heavy quark and
$\omega_{UV}$ is a cut-off isolating the narrow transport peak. It can
then be written as
\begin{equation}
\kappa = \frac{1}{3T} \sum_{i=1,3} \lim_{M \rightarrow \infty}
\frac{1}{\chi_{00}} \int dt d {\mathbf x} \langle \frac{1}{2}
\left( \left[ \phi^\dagger g E^i \phi - \theta^\dagger g E^i \theta
    \right] (t,{\mathbf x}), \left[ \phi^\dagger g E^i \phi -
    \theta^\dagger g E^i \theta \right] (0,{\mathbf 0}) \right)
\rangle ,
\label{kappa}
\end{equation}
where $\phi$ is the non-relativistic spinor for heavy quark, $\theta$
is that for heavy anti-quark, and $E^i$ is the color electric field.
\begin{figure}[ht]
\centering
\begin{minipage}[b]{0.43\linewidth}
\includegraphics[width=\textwidth]{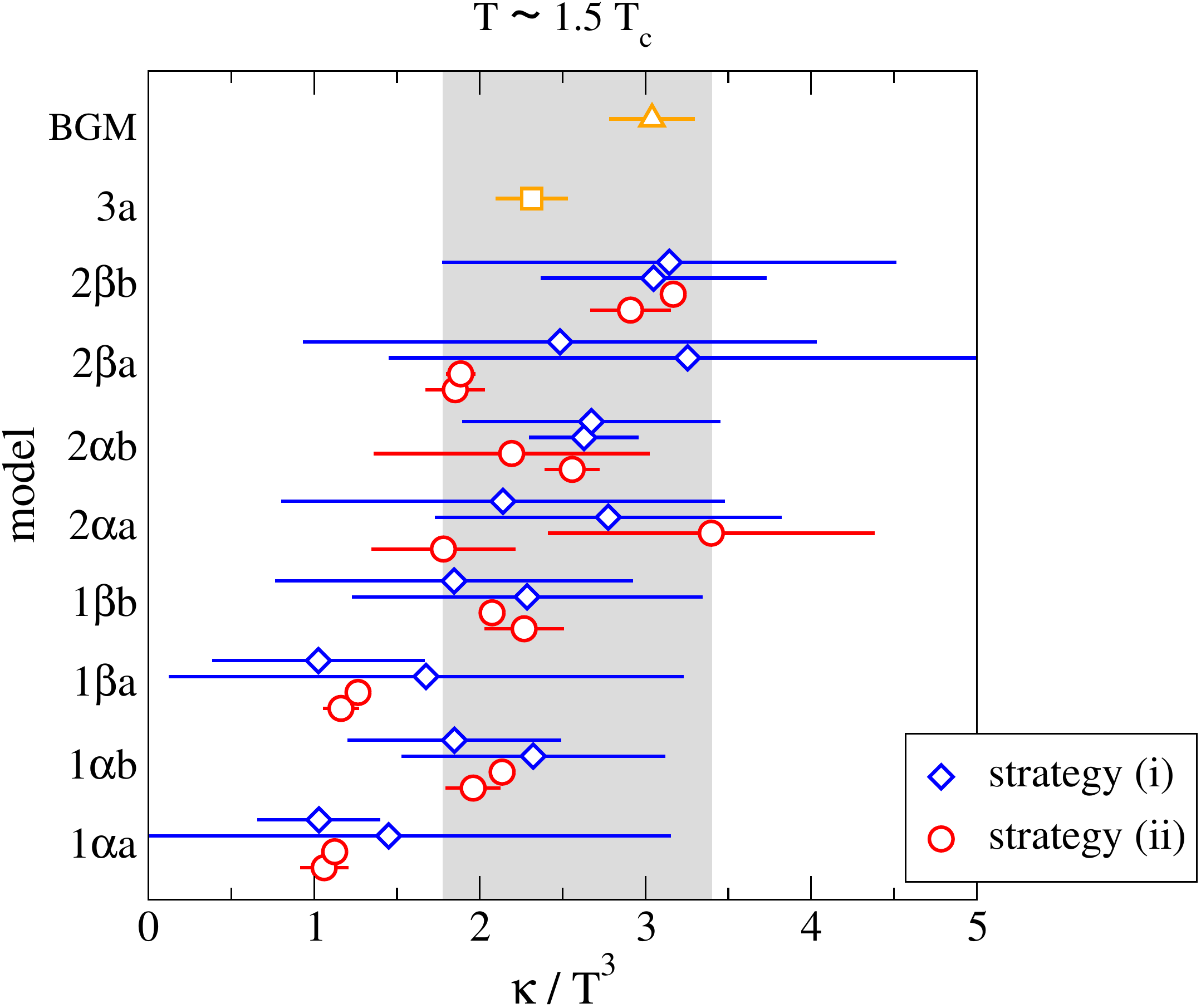}
\end{minipage}
\caption{The fitted results from various fitting strategies and
  fitting forms \cite{Francis:2015daa}, where the gray band is their
  final estimate (refer to \cite{Francis:2015daa} for the various symbols).} 
\label{kinetic}
\end{figure}
One can express Eq. \ref{kappa} as
\begin{equation}
G_E (\tau) = -\frac{1}{3} \sum_{1=1,3} \frac{ \langle {\rm Re} {\rm Tr} U
  (\beta, \tau) g E_i (\tau, {\mathbf 0}) U(\tau,0) g E_i (0,{\mathbf
  0}) \rangle} {\langle {\rm Re} {\rm Tr} \left[ U (\beta,0) \right]
  \rangle} ,
\label{kappa_E}
\end{equation}
where $U$ is the product of the time directional links. This is
amenable to a lattice calculation. It is still a difficult lattice
measurement because the observable is gluonic and suffers from large
statistical noise. \cite{Francis:2015daa} announced result of their
multi-year effort. On large quenched lattices ($64^3 \times 16 \sim
192^3 \times 48$), at a fixed temperature ($T = 1.5 T_c$), they
performed a continuum extrapolation of lattice-measured correlators
(Eq. \ref{kappa_E}) obtained with multi-level (actually two-level)
algorithm as the first step. Then, instead of applying a general
Bayesian reconstruction of the full spectral function for
Eq. \ref{kappa_E}, the IR limit and the UV limit of the spectral
function is argued to be a specific functional form. These two limits
are interpolated using various fitting forms and strategies. Fitted
results are also tested against standard Maximum Entropy
Method. Fig. \ref{kinetic} shows the fitted results of
$\kappa/T^3$. The gray band corresponds to $\kappa/T^3 = 1.8 - 3.4$,
which gives an estimate for the kinetic equilibriation time scale,
\begin{equation}
\tau_{\rm kin} = \frac{1}{\eta_D} = (1.8 \cdots 3.4)
\left(\frac{T_c}{T}\right)^2 \left(\frac{M}{1.5 {\rm GeV}} \right)
     {\rm fm/c} .
\end{equation}
This suggests that near $T_c$, charm quark kinetic equilibriation is
as fast as light parton kinetic equilibriation which is $\sim 1$ fm/c.

\subsection{Chemical equilibriation}

Similar to the kinetic equilibriation of heavy quarks in QGP, one can
define the chemical equilibriation rate as a transport coefficient
from a spectral function of a density-density correlator
\cite{Bodeker:2012gs, Bodeker:2012zm}. The chemical equilibriation is
related to how the number density ($n$) is adjusted, where
\begin{equation}
(\partial_t + 3h) n = - c (n^2 - n_{\rm eq}^2)
\end{equation}
in Boltzmann equation approach and
\begin{equation}
(\partial_t + 3h) n = - \Gamma_{\rm chem} (n - n_{\rm eq}) + {\cal O}
  (n - n_{\rm eq})^2
\end{equation}
in a linearized form near the equilibrium where $h$ is a kind of
``Hubble expansion constant'' if expanding QGP is
considered. $\Gamma_{\rm chem} = 2 c n_{\rm eq}$ is called the
chemical equilibriation rate. The linearized form can be described in
terms of a Langevin equation
\begin{equation}
\frac{\partial }{\partial t} \delta n(t) = - \Gamma_{\rm chem} \delta n(t)
+ \xi (t), \;\;\;\;\;\; \langle \langle \xi (t) \xi (t')\rangle
\rangle = \Omega_{\rm chem} \delta (t-t'), \;\;\;\;\; \langle
\langle \xi (t) \rangle \rangle = 0 
\end{equation}
where $\delta n (t) = n - n_{\rm eq}$ and $\langle \langle \cdots \rangle
\rangle$ denotes the average over the Langevin noise. A transport
property can be studied through the spectral function of a two-point
correlator 
\begin{equation}
\Delta (t,t') = \langle \frac{1}{2} \{ \delta n (t), \delta n (t') \}
\rangle .
\end{equation}
$\Omega_{\rm chem}$ and $\Gamma_{\rm chem}$ is related to the tail of
the spectral function for $\Delta (t,t')$,
\begin{equation}
\Omega_{\rm chem} = \lim_{\Gamma_{\rm chem} \ll \omega \ll
  \omega_{UV}} \omega^2 \int_{-\infty}^{\infty} dt \; e^{i\omega
  (t-t')} \langle \frac{1}{2} \{\delta n(t), \delta n(t') \} \rangle,
\;\;\;\; \Gamma_{\rm chem} = \frac{\Omega_{\rm chem}}{2 \chi_f M^2},
\label{chemical}
\end{equation}
where $\chi_f$ is the quark-number susceptibility related to the heavy
flavour. For a lattice calculation of the chemical equilibriation of
heavy quark, we need the imaginary-time formulation of
Eq. \ref{chemical}. Since in non-relativistic limit,
\begin{equation}
H = M (\theta^\dagger \theta - \phi^\dagger \phi),
\end{equation}
is related to the heavy quark number operator, the density-density
correlator in the imaginary time formalism,
\begin{equation}
\Delta (\tau) = \int d^3 {\mathbf x} \; \langle H (\tau, {\mathbf x})
H (0, {\mathbf 0}) \rangle , \;\;\; 0 < \tau < \frac{1}{T} = \beta ,
\end{equation}
is considered. Within NRQCD framework, the first order perturbation
due to pair annihilation for S-wave (${\rm Im} f_1 (^1 S_0)
\theta^\dagger \phi \theta \phi^\dagger$) gives
\begin{equation}
\Delta (\tau) \approx \frac{{\rm Im} f_1 (^1 S_0)}{\pi M^2} \int d^3
       {\mathbf x} \int d^3 {\mathbf y} \int_0^\beta d \tau_1
       \int_0^\beta d \tau_2  \frac{ \langle H (\tau, {\mathbf x}) H
         (0, {\mathbf 0}) (\theta^\dagger \phi) (\tau_1, {\mathbf 0})
         (\phi^\dagger \theta (\tau_2, {\mathbf 0}) \rangle }{|\tau_1
         - \tau_2 |} .
\label{NRQCDdelta}
\end{equation}
Then, from the tail of the spectral function,
\begin{equation}
\Omega_{\rm chem} = 16 {\rm Im} f_1 (^1 S_0) \frac{1}{Z} \sum_{m, n}
e^{-\beta E_m} \langle q \overline{q} m | \theta^\dagger \phi |n
\rangle \langle n | \phi^\dagger \theta | q \overline {q} m \rangle =
16 {\rm Im} f_1 (^1 S_0) \frac{1}{Z} {\rm Tr} \left[ e^{-\beta {\cal
      H}} (\theta^\dagger \phi) (0^+, {\mathbf 0}) (\phi^\dagger
  \theta) (0, {\mathbf 0}) \right], 
\end{equation}
where $q, \overline{q}$ are heavy quark and heavy anti-quark, and
$m,n$ denote the other degrees of freedom. Note that
\begin{equation}
\frac{1}{Z} {\rm Tr} \left[ e^{-\beta {\cal
      H}} (\theta^\dagger \phi) (0^+, {\mathbf 0}) (\phi^\dagger
  \theta) (0, {\mathbf 0}) \right] = \langle (\theta^\dagger \phi)
(0^+, {\mathbf 0}) (\phi^\dagger \theta) (0, {\mathbf 0}) \rangle =
     {\rm Tr} \langle G^\theta (\beta, {\mathbf 0}; 0, {\mathbf 0})
     G^{\theta \dagger} (\beta, {\mathbf 0}; 0, {\mathbf 0}) \rangle 
\end{equation}
and
\begin{equation}
\chi_f = \int d^3 {\mathbf x} \langle (\theta^\dagger \theta + \phi
\phi^\dagger) (\tau, {\mathbf x}) (\theta^\dagger \theta + \phi
\phi^\dagger) (0, {\mathbf 0})\rangle = 2 {\rm Re} {\rm Tr} \langle
G^\theta (\beta, {\mathbf 0}; 0 {\mathbf 0}) \rangle .
\end{equation}
Then, the thermal averaged Sommerfeld factors become
\begin{equation}
\overline{S}_1 = \frac{P_2}{P_1^2}, \;\;\;\; \overline{S}_8 =
\frac{N_c^2 P_3 - P_2}{(N_c^2 - 1) P_1^2} 
\end{equation}
with
\begin{equation}
P_1 = \frac{1}{2N_c}{\rm Re} \langle G^\theta_{\alpha \alpha;ii}
(\beta, {\mathbf 0}; 0, {\mathbf 0}) \rangle , \;\;\;\;\; P_2 = \frac{1}{2N_c}
\langle G^\theta_{\alpha \sigma ;ij} (\beta, {\mathbf 0}; 0, {\mathbf
  0}) G^{\theta \dagger}_{\sigma \alpha ;ji} (\beta, {\mathbf 0}; 0,
        {\mathbf 0}) \rangle ,
\end{equation}
\begin{equation}
P_3 = \frac{1}{2N_c^2} \langle G^\theta_{\alpha \alpha ;ij}
(\beta, {\mathbf 0}; 0, {\mathbf 0}) G^{\theta \dagger}_{\sigma \sigma ;ji}
(\beta, {\mathbf 0}; 0, {\mathbf 0}) \rangle,
\end{equation}
for the color singlet channel and the color octet channel
respectively. The Sommerfeld factor ($S$) usually means the
enhancement due to the attractive long range interaction between
slowly moving, annihilating heavy particles (heavy quark and heavy
anti-quark for QCD) compared to the Born matrix  element for the
pair annihilation ($|{\cal M}_{\rm resummed} |^2 = S | {\cal M}_{\rm
  tree} |^2 $) \cite{Sommerfeld:1931}. Then, 
\begin{equation}
\Gamma_{\rm chem}^1 = \frac{{\rm Im} f_1 (^1S_0)}{M^2} \; \chi_f \;
\frac{\overline{S}_1}{N_c}, \;\;\;\;\;
\Gamma_{\rm chem}^8 = \frac{{\rm Im} f_8 (^1S_0)}{M^2} \; \chi_f \;
\frac{(N_c^2-1)\overline{S}_8}{2N_c^2} .
\end{equation}
or if Boltzmann equation is considered \cite{Bodeker:2012gs},
\begin{equation}
 \Gamma^{ }_{\rm {chem}}
 \; \simeq \;
 \frac{8\pi\alpha_s^2}{3 M^2}
    \left( \frac{ M T } { 2 \pi  } \right) ^{ 3/2 } e ^{ - M/T } 
 \left[
   \frac{\bar{S}^{ }_1}{3} + 
 \left( \frac{5}{6} + Nf \right) \bar{S}^{ }_8 
 \right]
\label{Eq:BoltzmannEstimate}
\end{equation}
with the assumption that the octet Sommerfeld factors are
spin-independent. 

\begin{figure}[ht]
\centering
\begin{minipage}[b]{0.35\linewidth}
\includegraphics[width=\textwidth]{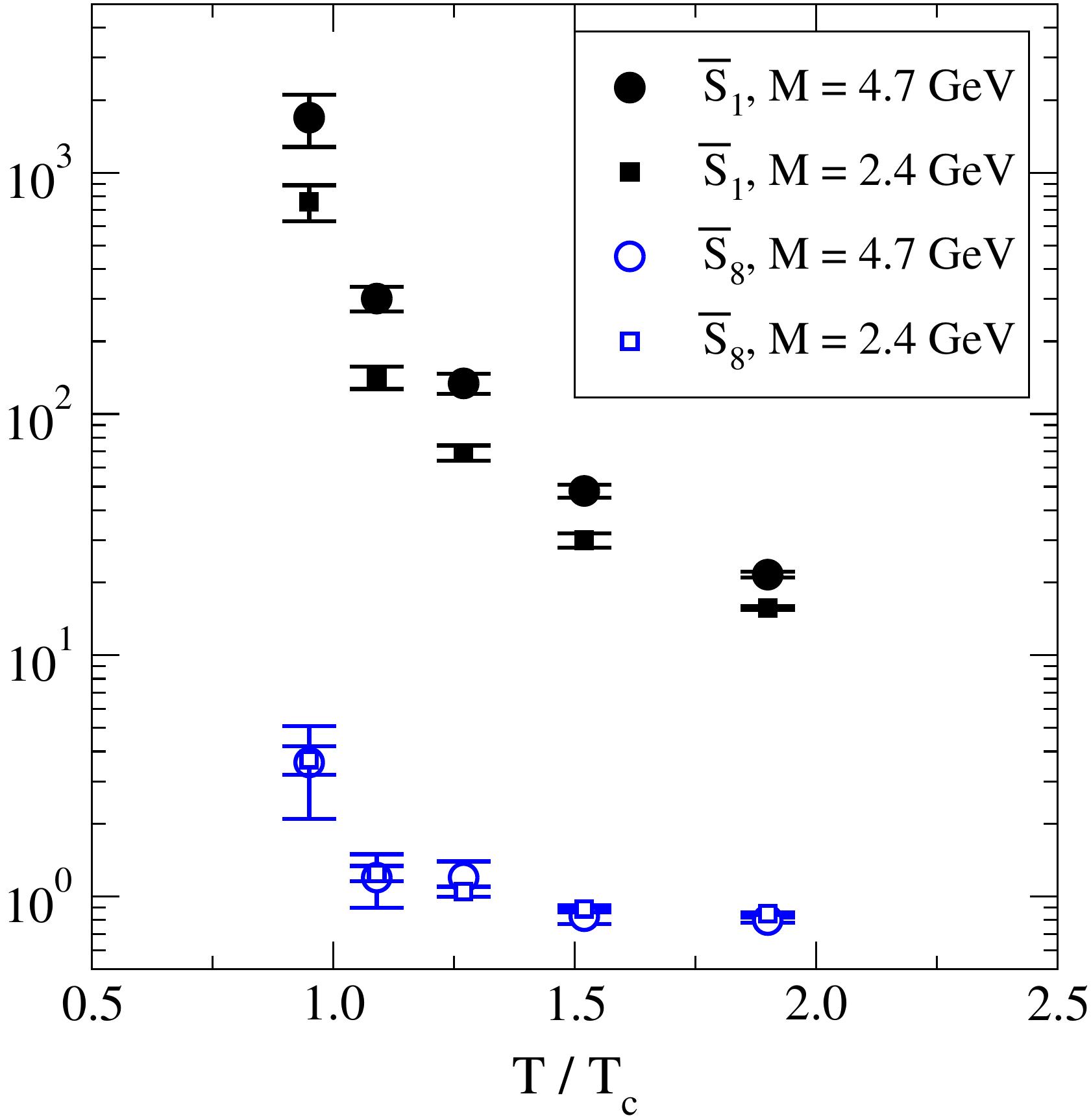}
\end{minipage}
\quad
\begin{minipage}[b]{0.35\linewidth}
\includegraphics[width=\textwidth]{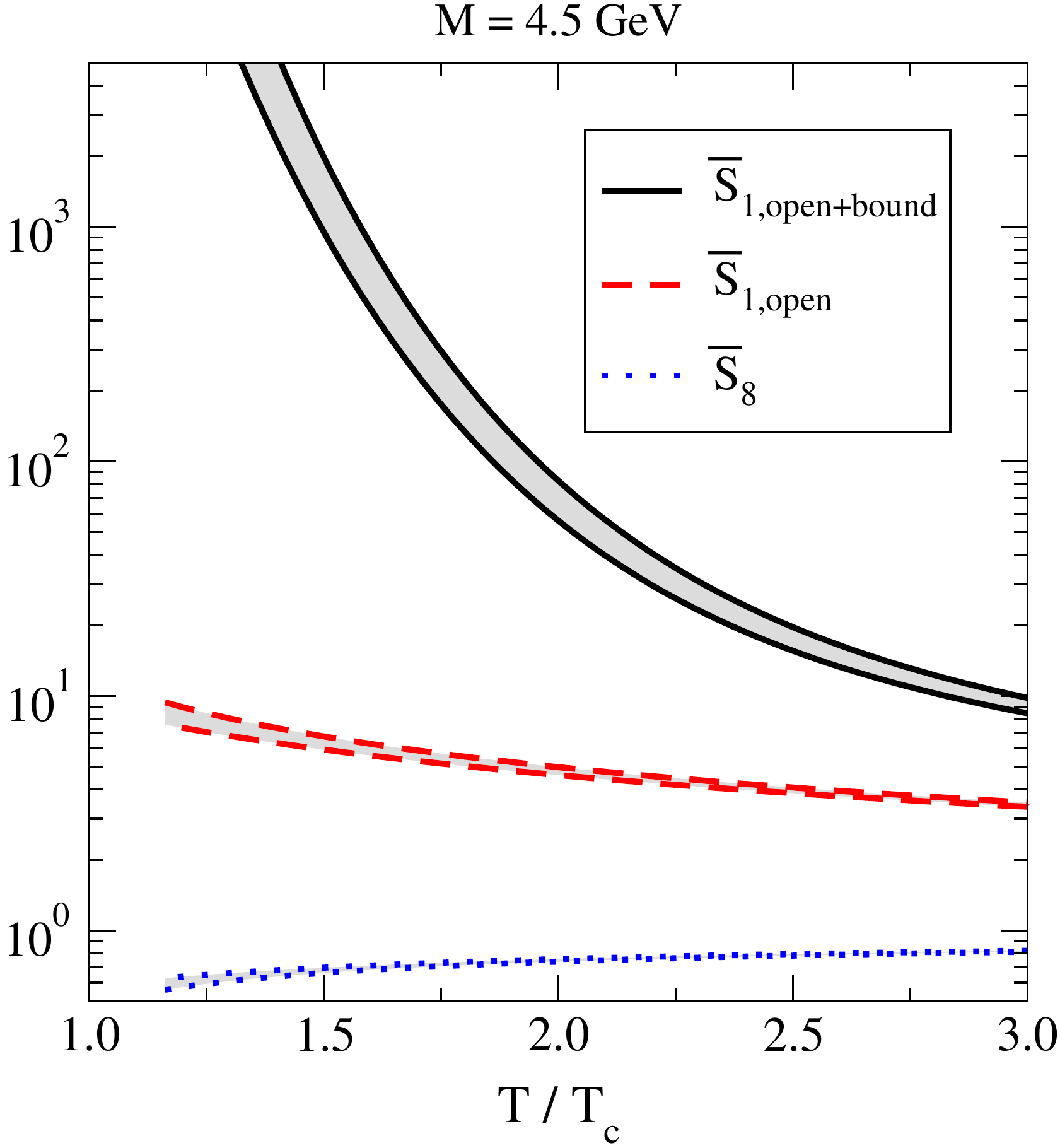}
\end{minipage}
\caption{The thermal averaged singlet channel Sommerfeld factor
  ($\overline{S}_1$) and octet channel factor ($\overline{S}_8$)
  \cite{Kim:2016zyy} (left). Note that the vertical axis is in the
  logarithmic scale. The right shows a perturbative calculation of
  these quantities using a spectral function with a bound state
  (black) and using a spectral function without a bound state (red)
  for the singlet channel}
\label{fig:chemicalEq}
\end{figure}

Using FASTSUM configurations \cite{Aarts:2014cda} ($24^3 \times
N_\tau, a_s/a_\tau = 3.5$ $a$-fixed ($a_s = 0.1227(8)$ fm), $N_f =
2+1$ with $m_\pi \simeq 400 $MeV, $m_K \simeq 500$ MeV),
\cite{Kim:2016zyy} calculated $\overline{S}_1$ and $\overline{S}_8$
(Fig. \ref{fig:chemicalEq}). A naive perturbative estimate for the
singlet Sommerfeld factor, $\overline{S}_1$, (the right figure, the
red line) $\sim {\cal O} (10)$. The lattice value ranges $\sim {\cal
  O} (100 - 1000)$. Only when a bound state contribution is included
in the model spectral function of heavy quark/heavy anti-quark system,
the analytic estimate becomes close to the lattice value (the right
figure, the black line). For the repulsive octet channel,
$\overline{S}_8$, both the lattice value and the analytic estimate
stays $\sim {\cal O} (1)$. The huge increase in the singlet Sommerfeld
factor illustrates the importance of non-perturbative effect near
$T_c$.
For a phenomenologically interesting case, one can estimate the
chemical equilibriation rate for a charm quark ($M \sim 1.5$ GeV)
using Eq. \ref{Eq:BoltzmannEstimate} at $T \sim 400$ MeV, with
$\overline{S}_1 \sim 15, \overline{S}_8 \sim 0.8$, which is
$\Gamma_{\rm chem}^{-1} \sim 150$ fm/c. This suggests that within the
QGP phase lifetime $\sim 10$ fm/c of the current relativistic heavy
ion collision, charm quark does not chemical equilibrate, as
expected.

\section{Quarkonium}

Despite long and intense efforts by the lattice community, a
quantitative understanding on whether bound states of heavy quark and
heavy anti-quark pair can exist in thermal medium based on first
principle calculation is still lacking. An early (qualitative) picture
of quarkonium ``melting'' based on a screened potential
\cite{Matsui:1986dk} is too simplistic since there is an imaginary part
of potential in thermal medium \cite{Laine:2006ns}. The recent
directions taken to study ``quarkonium in medium'' are roughly
categorized as,
\begin{itemize}
\item (1) calculate the potential between heavy quarks from the spectral
  function of Wilson loop/Wilson line correlator using $T \neq 0$
  gauge fields (see e.g., \cite{Rothkopf:2011db, Burnier:2012az,
    Burnier:2014ssa}). Or (2) calculate point-split quarkonium
  correlators of relativistic heavy quark propagators using $T \neq 0$
  gauge fields and define the potential by ``HAL QCD method''
  \cite{Ikeda:2011bs} (e.g, \cite{Iida:2011za, Allton:2015ora}). Then
  solve the Schr\"odinger equation for the quarkonium states using the
  lattice calculated potential \cite{Burnier:2015tda,Burnier:2016kqm}.
\item calculate a fully relativistic heavy quark propagator using $T
  \neq 0$ gauge fields and obtain quarkonium correlators from this
  relativistic heavy quark propagator. Then, obtain the spectral
  function of the temporal quarkonium correlator using Bayesian method
  \cite{Aarts:2007pk}. Full spectral features including a transport
  peak, bound state features and melting of bound states are
  expected. 
\item calculate heavy quark propagator using NRQCD under the
  background of $T \neq 0$ gauge fields and obtain non-relativistic
  temporal quarkonium correlator. Then, obtain the spectral function
  of the quarkonium correlators using various Bayesian methods
  \cite{Aarts:2010ek, Aarts:2011sm, Aarts:2012ka, Aarts:2013kaa,
    Aarts:2014cda, Kim:2014iga}. In NRQCD, absence of a transport
  feature in the spectral function (since $\sim 2M$ scale is
  integrated out) allows us to focus on the binding and melting
  features of states in the spectral function. 
\end{itemize}

Here we concentrate on the results from lattice NRQCD study of
quarkonium in non-zero temperature, i.e., studies of $T \neq 0$
quarkonium correlators ($G(\tau/a_\tau), \tau/a_\tau = 0, \cdots (N_\tau -
1)$, $a_\tau$ is the temporal lattice spacing) by calculating a heavy
quark propagator using lattice NRQCD and obtain the spectral functions
of the NRQCD quarkonium correlators using Bayesian method. There are
many practical advantages of NRQCD formalism over full QCD for a
lattice study of quarkonium: quarkonium correlators from a heavy quark
propagator which is calculated with NRQCD is highly accurate
(typically the statistical error is $\sim {\cal O} (10^{-4})$)
compared to the quarkonium correlators calculated with relativistic
QCD. NRQCD heavy quark propagator can be calculated fast since NRQCD
Lagrangian is a first order in time and is numerically an initial
value problem which allows a larger $\tau$ than a relativistic
propagator. However, the continuum limit of the correlators can not be
taken because the spatial lattice spacing must satisfy $M a_s \sim
{\cal O} (1) $.

Given $G (\tau)$, 
\begin{equation}
G(\tau) = \int_0^\infty \frac{d \omega}{2\pi} K (\omega, \tau)
\rho (\omega) \;\;\;\;\ 0 \leq \tau < \frac{1}{T}
\label{spectralft}
\end{equation}
where the kernel $K(\tau, \omega)$ becomes $\sim e^{-\omega \tau}$ for
NRQCD and $(e^{-\omega \tau} + e^{-\omega (1/T - \tau)})/(1 -
e^{-\omega T}) = \cosh (\omega (\tau - 2/T)) / \sinh (\omega 1/2T)$
for full QCD, non-zero temperature behavior is studied by the
temperature dependence of the spectral function, $\rho (\omega)$. The
integral transform, Eq. \ref{spectralft}, shows the crux of studying
the in-medium property of quarkonium through the spectral function of
the Euclidean correlator. Since we do not know the analytic structure
of the spectral function, this inverse integral transform problem is
ill-posed. To make the numerical problem worse, the number of temporal
lattice sites for $G(\tau)$ at $T \neq 0$ is typically smaller than
that at zero temperature while the spectral structure is expected to
be more complicated at $T \neq 0$ than that at zero temperature (i.e.,
we need to extract more ``information'' with less
``data''). Simplified form of the kernel for NRQCD gives several
advantages over the kernel for full QCD: (1)a ``constant contribution
problem'' is absent \cite{Umeda:2007hy, Aarts:2002cc,
  Petreczky:2008px}. (2)$\tau/a_\tau$ of $G(\tau/a_\tau)$ can extend its range
to $1/T a_\tau - 1$. (3)the inverse integral transform becomes the
inverse Laplace transform. For a given lattice $G(\tau/a_\tau)$ with finite
error bars, there still exist numerous $\rho (\omega)$'s which satisfy
Eq. \ref{spectralft}. To this problem, Bayes theorem for the
conditional probability
\begin{equation}
P [X|Y] = P[Y|X] P[X] / P[Y] ,
\end{equation}
is applied where $P [X|Y]$ is the probability of observing the event X
given that the event Y is true. For a given Data $D$ and a prior
knowledge ($H$), the probability for the spectral function ($\rho$) is
\begin{equation}
P [\rho | D, H] \propto P[D|\rho,H] P[\rho|H], \;\;\; P[D|\rho,H] =
e^{-L}, \;\;\; L = \frac{1}{2} \sum_{i} (D_i - D_i^\rho)^2/\sigma_i^2
, \;\;\; P[\rho | H] = e^{-S}, \;\;\; S = S[\rho(\omega), m(\omega)], 
\end{equation}
where
\begin{equation}
P[\rho | H] = e^{-S}, \;\;\; S = S[\rho(\omega), m(\omega)] .
\end{equation}
Here $S$ is a prior and $m (\omega)$ is a default model. Currently,
two different priors, Shannon-Jaynes entropy for $S$
\begin{equation}
S_{SJ} = \alpha \int d \omega \; \left(\rho - m - \rho \; \log
(\frac{\rho}{m})\right) 
\end{equation}
which is called ``Maximum Entropy Method \cite{Asakawa:2000tr} and new
prior \cite{Burnier:2013nla} 
\begin{equation}
S_{BR} = \alpha \int d \omega \; \left(1 - \frac{\rho}{m} + \log
(\frac{\rho}{m})\right) 
\end{equation}
are under studies. Since only with infinite number of data points in
$G(\tau)$ and zero statistical errors all methods should agree 
\cite{Asakawa:2000tr}, many different priors must be tested with
finite number of data points and non-zero statistical errors. 

\subsection{Upto last year's conference} 

\begin{figure}[ht]
\centering
\begin{minipage}[b]{0.48\linewidth}
\includegraphics[width=\textwidth]{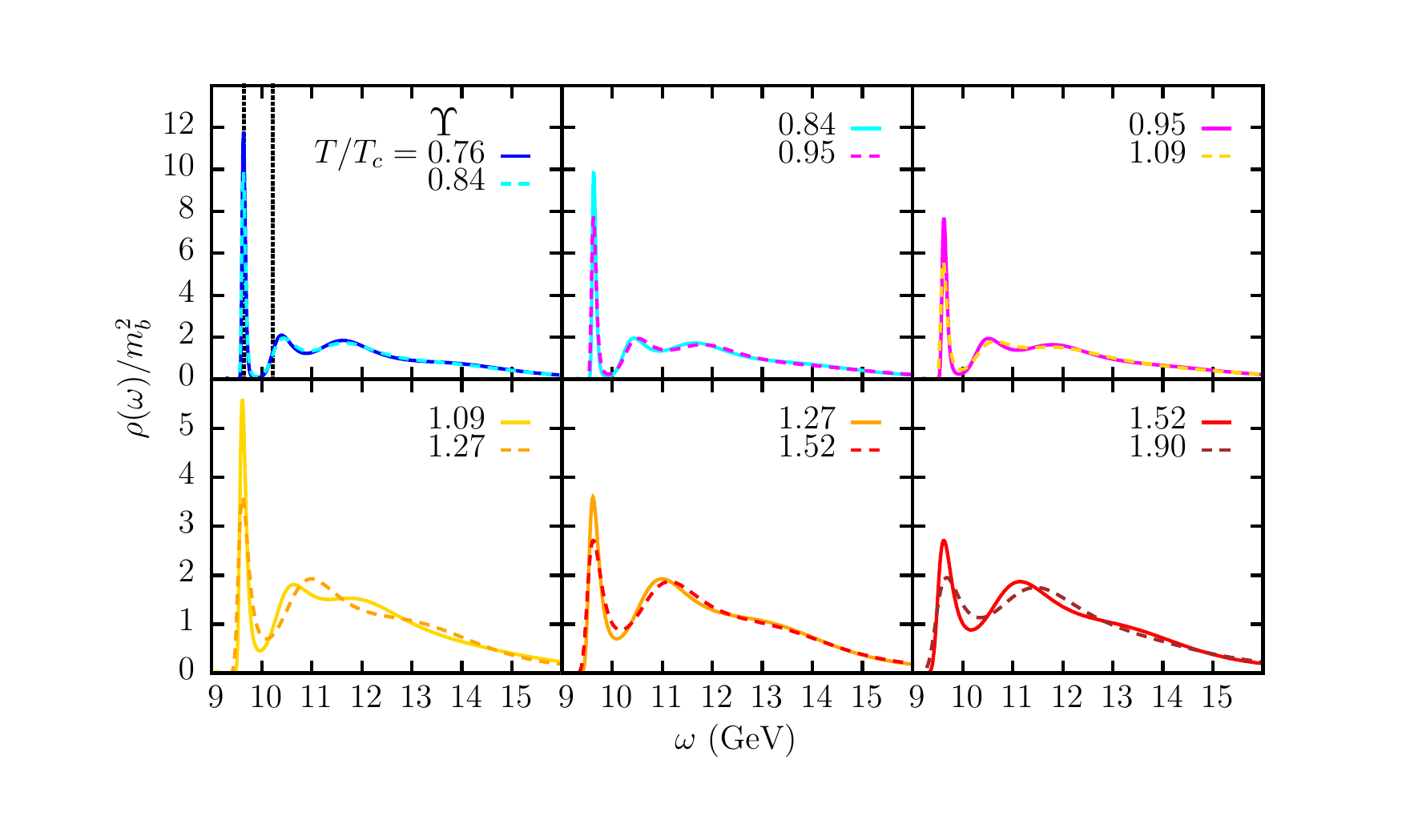} 
\end{minipage}
\quad
\begin{minipage}[b]{0.48\linewidth}
\includegraphics[width=\textwidth]{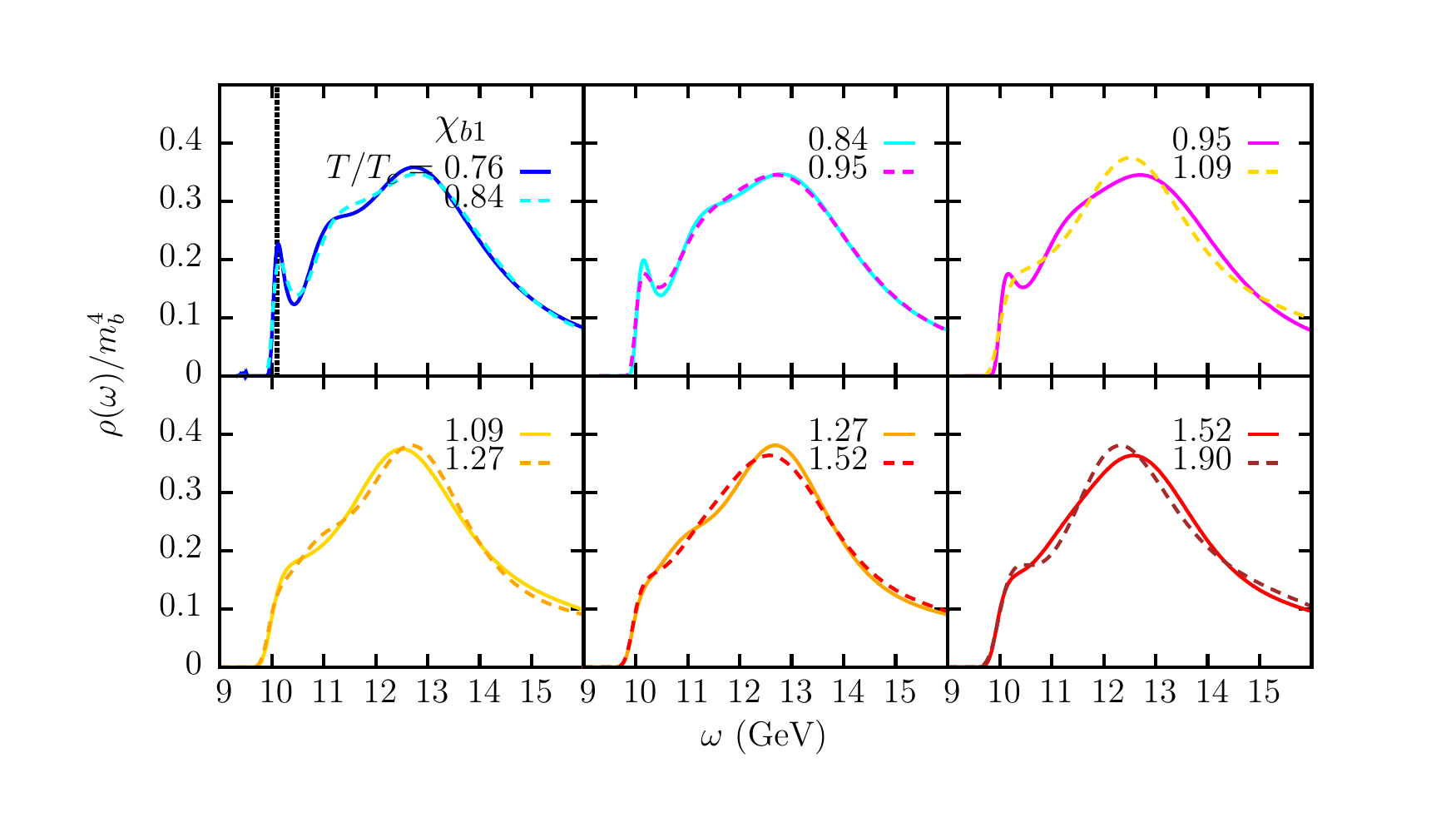}
\end{minipage}
\caption{Spectral functions for the $\Upsilon$ channel (left) and those
  for the $\chi_{b1}$ channel (right) at successive temperatures from
  FASTSUM collaboration using NRQCD heavy quark propagator and MEM for
  the reconstruction of spectral function \cite{Aarts:2014cda}.} 
\label{fig:FASTSUM}
\end{figure}

Using anisotropic lattices with a fixed lattice scale and temperature
change by $N_\tau$, FASTSUM collaboration calculated $T \neq 0$
quarkonium correlator using NRQCD and obtained the spectral functions
of S-wave and P-wave quarkonium correlators using MEM. Using the 1st
generation configurations ($12^3 \times N_\tau, a_s/a_\tau = 6.0, N_f
= 2$ where $m_\pi/m_\rho \approx 0.54, T_c \approx 210 {\rm MeV}$
using Two-plaquette Symanzik Improved gauge action and the
fine-Wilson, coarse-Hamber-Wu fermion action with stout-link smearing
\cite{Aarts:2007pk}), FASTSUM collaboration calculated $T \neq 0$
quarkonium correlator \cite{Aarts:2010ek} using NRQCD and obtained the
spectral functions of quarkonium correlators using MEM for S-wave
channel \cite{Aarts:2011sm}, moving S-wave channel
\cite{Aarts:2012ka}, and P-wave channel
\cite{Aarts:2013kaa}. Anisotropic lattices allow a larger number of
temporal lattice sites while satisfying $M a_s \geq 1$ (for the
validity of NRQCD). With a fixed $a_\tau$, the zero point energy shift
associated with NRQCD formalism needs to be fixed only once.

\begin{figure}[ht]
\centering
\begin{minipage}[b]{0.43\linewidth}
\includegraphics[width=0.76\textwidth,angle=-90]{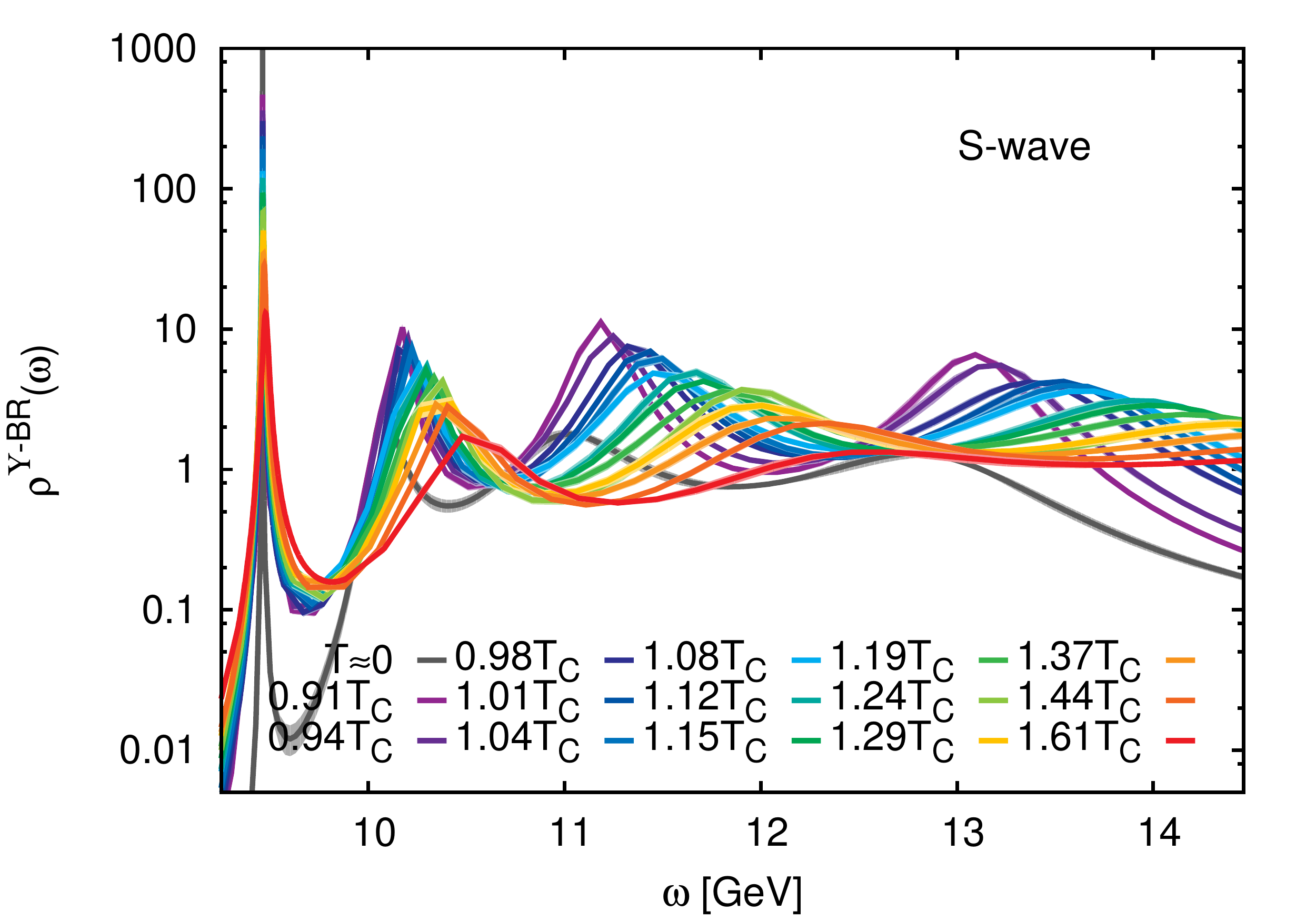}
\end{minipage}
\hspace{0.75cm}
\quad
\begin{minipage}[b]{0.43\linewidth}
\includegraphics[width=0.75\textwidth,angle=-90]{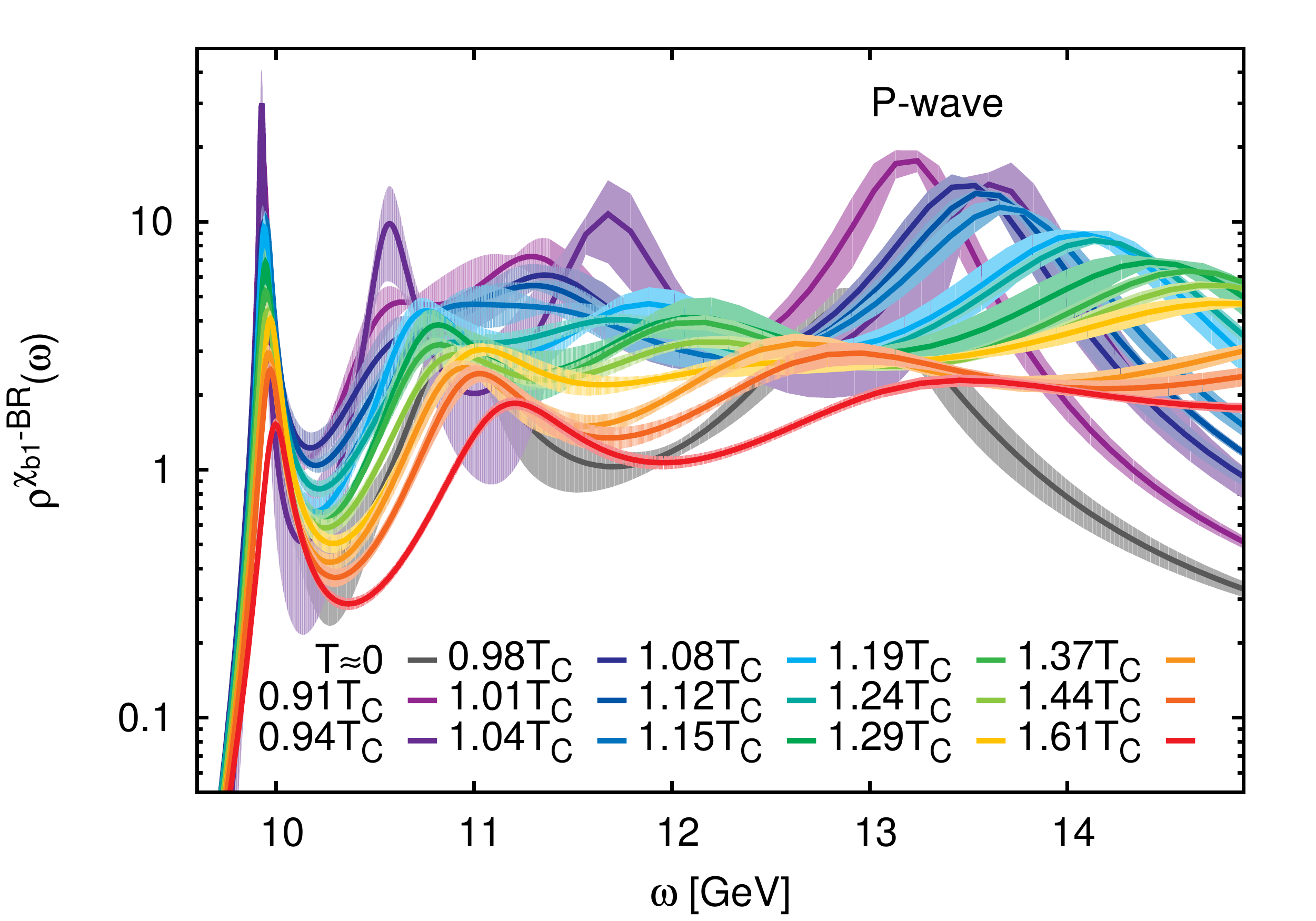}
\end{minipage}
\caption{Spectral functions for the $\Upsilon$ channel (left) and
  those for the $\chi_{b1}$ channel at successive temperature from
  \cite{Kim:2014iga} using NRQCD heavy quark propagator and new
  Bayesian prior \cite{Burnier:2013nla} for the reconstruction of
  spectral function.}
\label{fig:KPR}
\end{figure}

From the spectral function of $\Upsilon$ channel, it is found that
$\Upsilon (1S)$ survives upto $T \sim 2.1 T_c$ while $\Upsilon (2S)$
and $\Upsilon (3S)$ are sequentially suppressed in agreement with
\cite{Chatrchyan:2012lxa}. $\chi_{b1}$ melts immediately above $T_c$
\cite{Aarts:2013kaa}. A detailed study of systematic errors is
performed: default-model dependence in the MEM prior, energy window
dependence in the spectral function reconstruction, dependence on the
number of configurations and dependence on the Euclidean time window
and found that the reconstructed spectral functions are
stable. However, with non-zero statistical errors in the data of
quarkonium correlators and finite number of the temporal lattice
sites, choice of the prior $S$ is not unique and reconstructed
spectral functions may depend on the choice. Using the same setup on
the 2nd generation configurations ($24^3 \times N_\tau, a_s/a_\tau =
3.5, N_f = 2+1$), FASTSUM collaboration confirmed their earlier
conclusions\cite{Aarts:2014cda}: the survival of $\Upsilon (1S)$ upto
$\sim 1.9 T_c$ and the melting of $\chi_{b1}$ just above
$T_c$. Fig. \ref{fig:FASTSUM} shows the spectral functions for the
$\Upsilon$ channel and that for the $\chi_{b1}$ channel.

For a quantitative understanding of the quarkonium melting, FASTSUM's
results need to be checked with a setup which has different
systematics. Authors of \cite{Kim:2014iga} calculated NRQCD heavy
quark propagator using a isotropic lattice configurations from HotQCD
($48^3 \times 12$, $N_f = 2+1$ light $m_\pi \sim 160$ MeV and $T_c =
154$ MeV) \cite{Bazavov:2011nk} where the temperature is changed by
changing the lattice spacing $a$ with fixed $N_\tau = 12$. The pro is
that the temperature can be changed continuously and a detailed
temperature scan is possible. The con is that the zero energy shift
for NRQCD needs to be fixed by accompanying $T = 0$ calculations.

Similar to the works by FASTSUM collaboration, a detailed study of
systematic errors is also performed. In addition, an investigation on
the prior dependence of the reconstructed spectral functions by using
two different priors, MEM and a new Bayesian prior
\cite{Burnier:2013nla}. \cite{Kim:2014iga} found that qualitatively
similar behavior for the $\Upsilon$ channel found by FASTSUM: the
survival of $\Upsilon (1S)$ upto the highest temperature they studied
($\sim 1.6 T_c$) from two different priors. The left figure of
Fig. \ref{fig:KPR} shows the spectral function of the $\Upsilon$
channel obtained with new Bayesian prior. However, qualitatively
different behaviors for the $\chi_{b1}$ channel is found: the spectral
function from MEM shows melting but the spectral function from new
Bayesian prior shows survival of $\chi_{b1}$ upto $\sim 1.6 T_c$ (the
right figure in Fig. \ref{fig:KPR}).

Reconstruction of spectral function with new Bayesian prior is
generally stronger in identifying peaks \cite{Burnier:2013wma} but is
susceptible to spurious ``ringing'' (i.e., reconstructs peaks for the
spectral function from lattice free quarkonium correlators although an
analytic result for the spectral function does not show such peaks
\cite{Kim:2014iga}). 

\subsection{Since last year's conference}

Bayesian reconstructions of spectral functions for the correlators
with finite data points and finite statistical errors are expected to
show prior dependence. Between MEM prior and new Bayesian prior, which
one is closer to the ``Bayesian continuum limit'' (infinite data
points and zero statistical errors) can only be tested by going
toward the Bayesian continuum limit. 

\begin{figure}[ht]
\centering
\begin{minipage}[b]{0.43\linewidth}
\includegraphics[width=\textwidth]{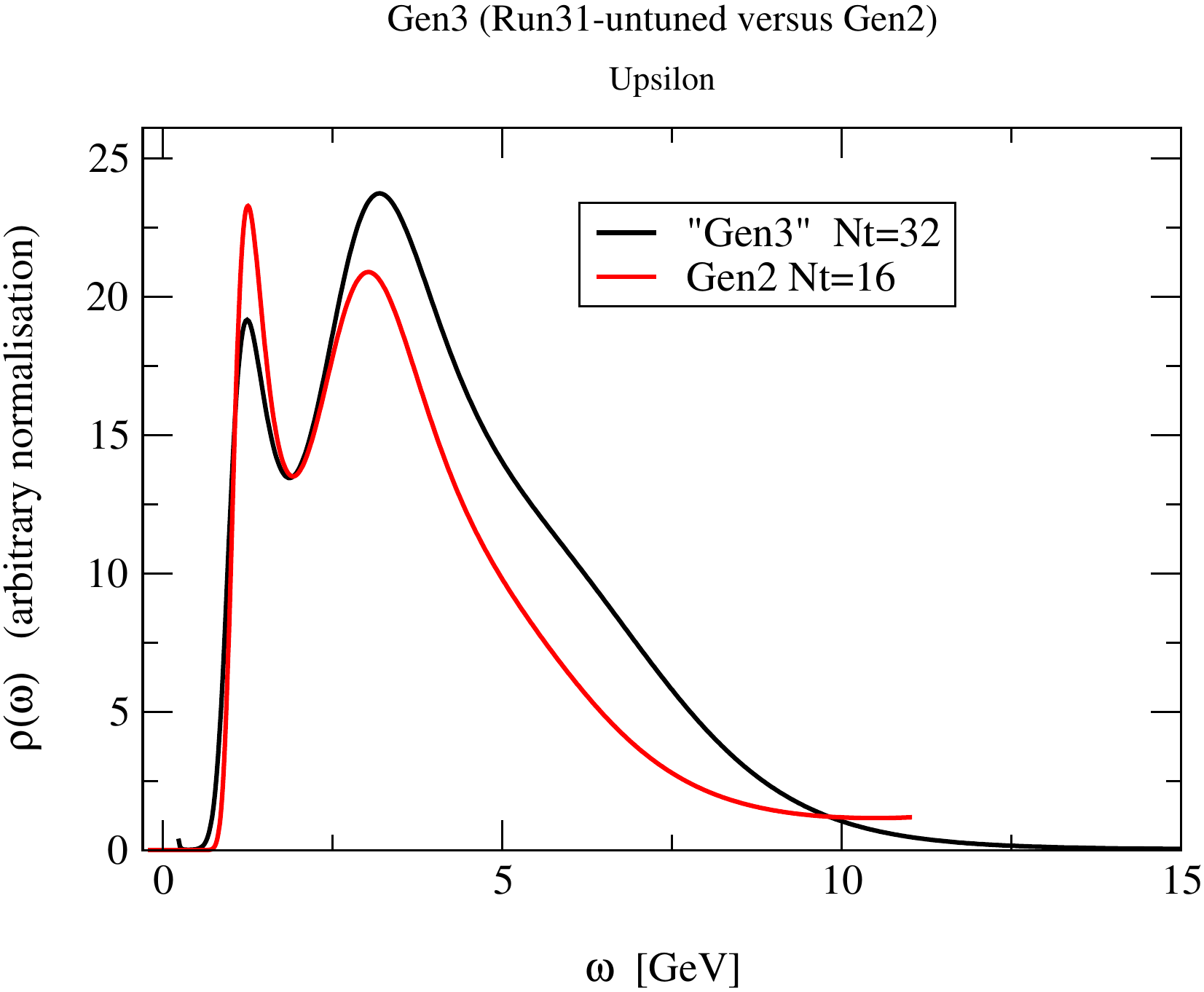}
\end{minipage}
\quad
\begin{minipage}[b]{0.45\linewidth}
\includegraphics[width=\textwidth]{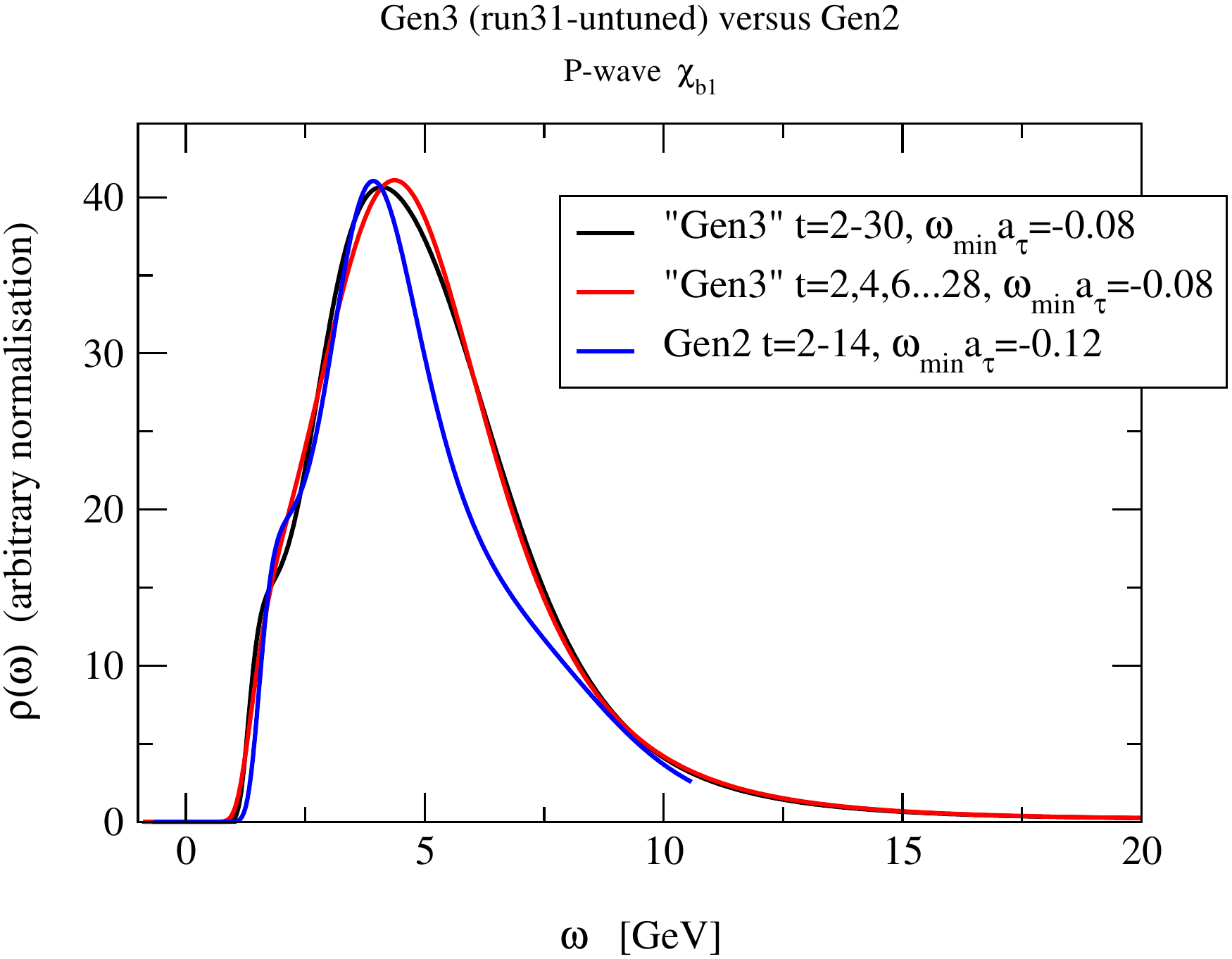}
\end{minipage}
\caption{Preliminary spectral functions for the $\Upsilon$ channel
  (left) and those for the $\chi_{b1}$ channel (right) from the 3rd
  Generation configurations (see the text for parameters).}
\label{fig:FASTSUM_prelim}
\end{figure}

\begin{figure}[ht]
\centering
\begin{minipage}[b]{0.43\linewidth}
\includegraphics[width=\textwidth]{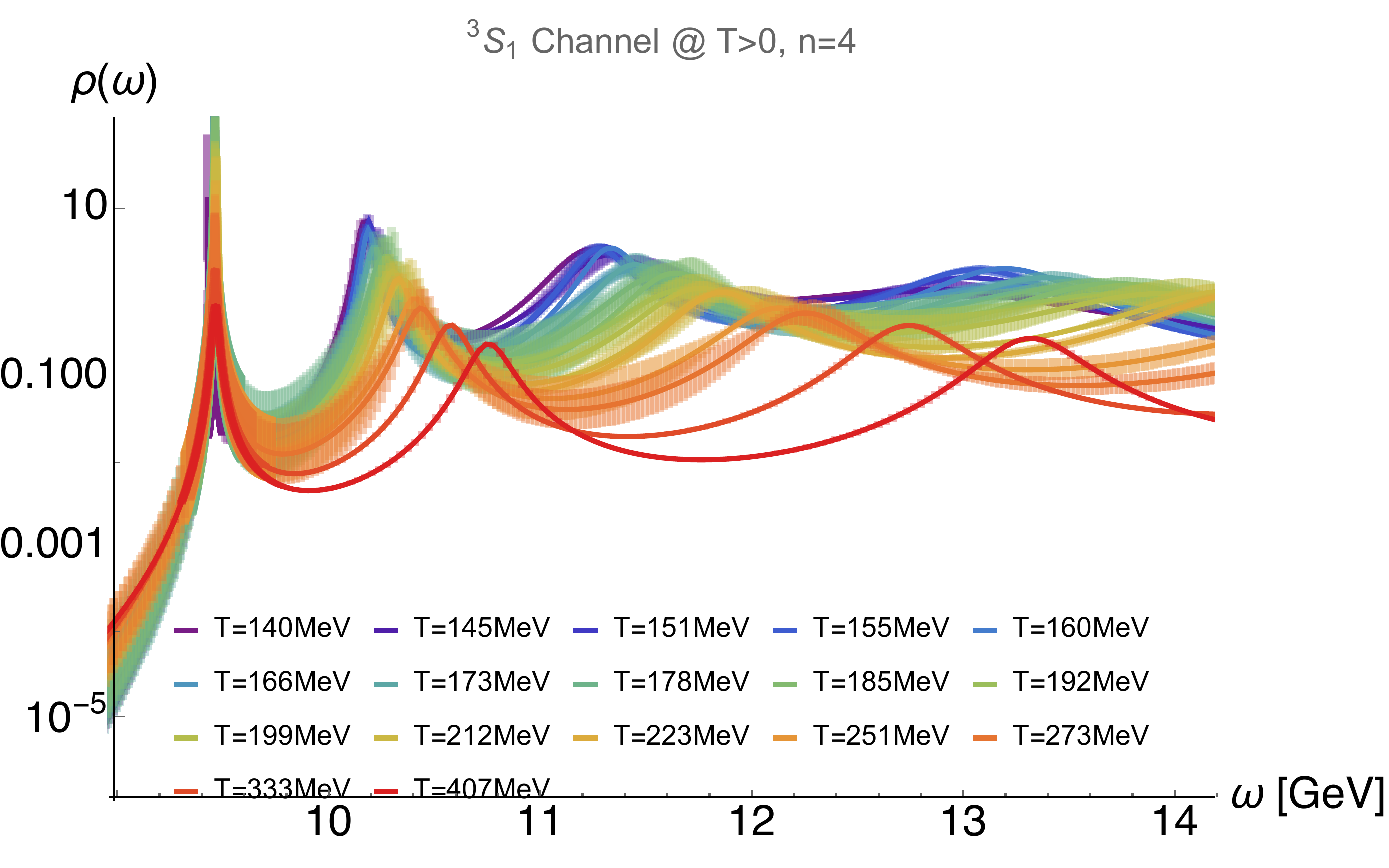}
\end{minipage}
\quad
\hspace{0.75cm}
\begin{minipage}[b]{0.45\linewidth}
\includegraphics[width=\textwidth]{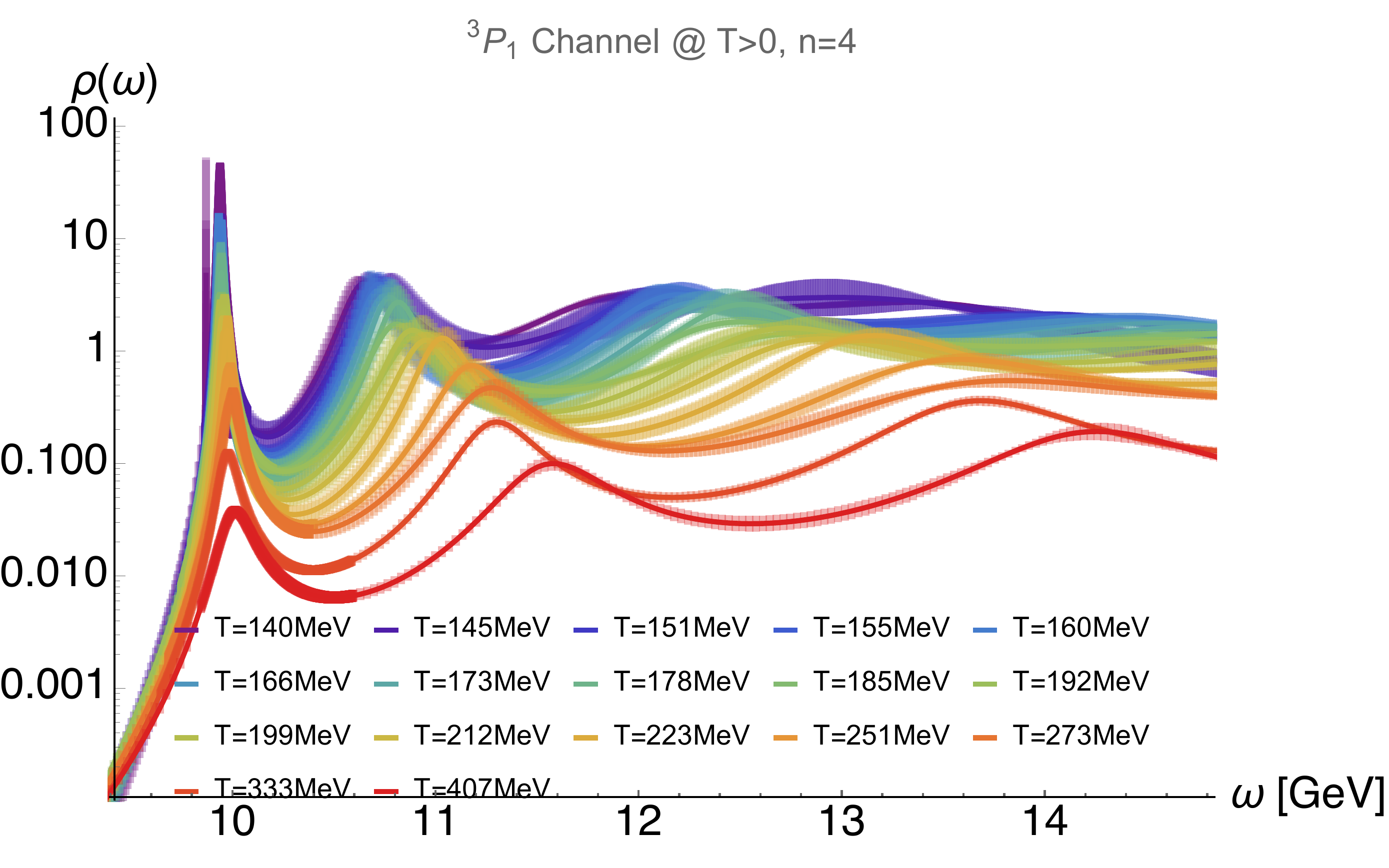}
\end{minipage}
\caption{Preliminary spectral functions for the $\Upsilon$ channel
  (left) and those for the $\chi_{b1}$ channel (right) from
  S.K. A. Rothkopf, P. Petreczky with four times higher statistics
  correlators and higher temperature (see the text for parameters).}
\label{fig:KPR_prelim}
\end{figure}

For the further study, FASTSUM collaboration focus on testing a
direction of the Bayesian continuum limit by halving the lattice
spacing (i.e., increasing the temporal lattice data points by twice)
while keeping other conditions the same. In
Fig.~\ref{fig:FASTSUM_prelim}, preliminary results of the spectral
functions from the ``3rd generation configurations ($32^3 \times
N_\tau, a_s/a_\tau = 6.85, N_f = 2+1$) show no significant differences
between the results from the 2nd generation configuration and those
from the preliminary 3rd generation configurations. On the other hand,
authors of \cite{Kim:2014iga} concentrate on increasing statistics
($\sim 4000$ correlators) so that the error bars in the correlators
gets small. Also, they explore bottomonium at higher temperatures ($T
= 273 (1.84T_c), 333(2.25T_c), 407 (2.75T_c)$ using new HotQCD gauge
configurations\cite{Bazavov:2014pvz}. Fig.~\ref{fig:KPR_prelim} show
preliminary results of the reconstructed spectral functions of
bottomonium correlators at higher statistics and higher
temperature. Interestingly, at higher temperature, $\chi_{b1}$ finally
shows melting behavior.

\section{Summary}

Some of recent progresses in the lattice studies of heavy flavours in
thermal environment are discussed after looking over the opportunities
offered by studying the properties of QGP through heavy quark flavours
and the challenges faced in such lattice studies. Focuses in this talk
are on the transport phenomena of heavy quarks in thermal medium and
the spectral properties of quarkonia in thermal medium.

Various effective field theories for heavy quark and quarkonia on a $T
= 0$ lattice are well understood and give us many accurate results on
the properties of QCD. The same effective field theory on a non-zero
temperature lattice alleviates the problem associated with various
scales in a heavy quark calculation (lattice spacing, Compton wave
length of heavy quark, and large physical size of a lattice) by
``integrating out'' momentum scale greater than the heavy quark mass.

In calculating transport properties of heavy quark such as the kinetic
equilibriation and the chemical equilibriation, heavy quark mass limit
offers a formulation of transport properties as a tail of the spectral
function, instead of the usual transport peak of the spectral function
as a transport coefficient. Since a transport peak of the heavy quark
spectral function is narrow ($\sim \alpha_s^2 T^2/M $), this
alternative formulation avoids the numerical difficulty of accessing a
narrow peak in the study of kinetic equilibriation and also new
observable defined in the formulation gives less complicated spectral
functions. It gives us a robust determination of the momentum
diffusion coefficient for heavy quark at least in the quenched
approximation. $\kappa/T^3 = 1.8 - 3.4$ for charm quark of
\cite{Francis:2015daa} gives ${\cal O} (1)$ fm/c as the kinetic
equilibriation time scale which is similar to the light parton kinetic
equilibriation time scale. Although it is difficult to generalize the
same setup to the case of fully dynamical lattices (because the new
observable is gluonic), quenched non-perturbative estimate of the
momentum diffusion constant is a significant step toward understanding
fast kinetic equilibriation of charm quark in relativistic heavy ion
collisions.

In the case of heavy quark chemical equilibriation, ``tail of a
spectral function as a transport coefficient'' allows a
non-perturbative definition of the thermal averaged Sommerfeld factor
(and related $\Gamma_{\rm chem}$) as a ratio of two-point correlators.
This factor is calculable using lattice NRQCD method (and the chemical
equilibriation rate, given as the heavy quark number susceptibility
times the Sommerfeld factor). Using dynamical $N_f=2+1$ lattices
($m_\pi \simeq 400 $MeV, $m_K \simeq 500$ MeV), \cite{Kim:2016zyy}
found much larger Sommerfeld enhancement ($\sim {\cal O} (100)$) than
estimated by a perturbative method. Phenomenologically, at $T \sim 400$
MeV, for a charm quark mass $M \sim 1.5$ GeV, $\Gamma_{\rm chem}^{-1}
\sim 150$ fm/c, which suggests that within the lifetime $\sim 10$ fm/c
of QGP phase in the current relativistic heavy ion collision
experiment, the chemical equilibriation is not achieved. Since the
chemical equilibriation rate changes rapid with temperature, however,
the planned Future Circular Collider \cite{Dainese:2016gch} heavy ion
program may yield charm quark chemical equilibriation.

Over the years, lattice community studied thermal behavior of
quarkonium by various methods. Recently, using lattice NRQCD to
calculate a heavy quark propagator and using Bayesian method to obtain
spectral functions of temporal quarkonium correlators constructed from
this NRQCD heavy quark propagator overcame various difficulties 
identified in studying quarkonium on a lattice. Up until this year's
lattice conference, two groups which have used this method in
different lattice setup reported that the survival of $\Upsilon (1S)$
upto $\sim 1.9 T_c ({\rm or} 1.6 T_c)$ and sequential suppression of
higher $\Upsilon$ states as temperature increases. But the results on
$\chi_{b1}$ states differ: FASTSUM reports the melting of P-wave
immediately above $T_c$ but the result of \cite{Kim:2014iga} hints at
the survival of $\chi_{b1}$ upto $\sim 1.6 T_c$. In these studies,
quarkonium correlators themselves can be calculated to high
accuracy. However, since Bayesian methods can give a unique result
only in the Bayesian continuum limit, further studies are
needed. FASTSUM group focuses on increasing the data points of
temporal quarkonium correlators. Preliminary result shows that
doubling the number of the temporal lattice sites does not change
significantly reconstructed spectral functions of $\Upsilon$ and
$\chi_{b1}$. Authors of \cite{Kim:2014iga} focus on increasing
temperature and increasing statistics. Their preliminary result show
that at higher temperature, $\chi_{b1}$ melts and higher statistics
reduces jack-knife errors in the reconstructed spectral
functions. More studies by both groups are forthcoming. Also, further
investigations on the Bayesian reconstruction method for the spectral
functions are needed.

{\bf Acknowledgements}

I would like to acknowledge that many discussions with G. Aarts,
C. Allton, N. Brambilla, Y. Burnier, A. Francis, O. Kaczmarek,
M. Laine, M.P. Lombardo, P. Petreczky and A. Vairo were helpful to my
understanding on ``heavy flavors in non-zero temperature''. This work
is supported by the National Research Foundation of Korea under grant
No.\ 2015R1A2A2A01005916 funded by the Korean government (MEST).

\end{document}